\documentclass[journal]{IEEEtran}

\usepackage{cite}
\usepackage{amsmath}
\usepackage{pgfplots}
\usepackage{amssymb}
\usepackage{tikz}
\usepackage{tikzscale}
\usepackage[linesnumbered,ruled]{algorithm2e}
\usepackage{pgfplots}
\usepackage{caption}
\usepackage{subcaption}

\newcommand{\betab}[0]{\boldsymbol{\beta}}

\definecolor{darkblue}{rgb}{0.267, 0.447, 0.769}
\definecolor{blue2}{rgb}{0.39, 0.58, 0.93}

\onecolumn
\begin{document}

\title{A Fiber Measurement System with Approximate Deconvolution Based on the Analysis of Fault Clusters in Linearized Bregman Iterations}

\author{Yuneisy~E.~Garcia Guzman,~\IEEEmembership{Student Member,~IEEE,}
        Felipe~Calliari, Gustavo~C.~Amaral,~\IEEEmembership{Member,~IEEE,}
        and~Michael~Lunglmayr,~\IEEEmembership{Member,~IEEE}
\thanks{Y. E. Garcia Guzman and M. Lunglmayr are with the Institute of Signal Processing, Johannes Kepler University, 4040 Linz, Austria (e-mail:\{yuneisy.garcia\_guzman ; michael.lunglmayr\}@jku.at).}
\thanks{F. Calliari and G. C. Amaral are with the Center for Telecommunication Studies of the Pontifical Catholic University of Rio de Janeiro, 22451-090 Rio de Janeiro, Brazil, (e-mail:\{felipe.calliari ; gustavo\}@opto.cetuc.puc-rio.br).}
\thanks{Manuscript received --}}

\maketitle

\begin{abstract}
Automatic detection of faults in optical fibers is an active area of research that plays a significant role in the design of reliable and stable optical networks. A fiber measurement system that combines automated data acquisition and processing represents a disruptive impact in the management of optical fiber networks with fast and reliable event detection. It has been shown in the literature that the linearized Bregman iterations (LBI) algorithm and variations can be successfully used for processing and accurately identifying faults in a fiber profile. One of the factors that impact the performance of these algorithms is the degradation of spatial resolution, which is mainly caused by the appearance of fault clusters due to a reduced number of iterations. In this paper, a method is proposed based on an approximate deconvolution approach for increasing the spatial resolution, possible after a thorough analysis of fault clusters that appear in the algorithm's output. The effect of such approximate deconvolution is shown to extend beyond the improvement of spatial resolution, allowing for better performances to be reached at shorter processing times. An efficient hardware architecture that implements the approximate deconvolution, compatible with the hardware structure recently presented for the LBI algorithm, is also proposed and discussed.
\end{abstract}

\IEEEpeerreviewmaketitle

\section{Introduction}

The Linearized Bregman Iterations (LBI) algorithm has recently been applied to detect unknown trend breaks in the presence of noise with remarkable success, achieving better results than several other methods \cite{lunglmayr2018tim, MDL_method, ge2018efficient, calliari2020jwcn, weid2016jlt} both in terms of estimation performance and processing speeds. The application context, fiber fault detection, requires fast and precise identification of the positions and magnitudes of the trend breaks in a so-called fiber profile \cite{derickson1998fiber}. The fiber profile, in turn, is available due to a data acquisition system that measures, under linearity assumptions, the impulse response of the fiber; this role is, in general, delegated to an Optical Time-Domain Reflectometry-based measurement apparatus, or OTDR \cite{barnoski1977optical}. OTDR devices based on different techniques are available commercially, finding applications in, e.g., passive fiber sensors for large-scale structure monitoring \cite{lee2003review, barrias2016review}; but are also currently investigated in a research context \cite{wang2015long, hu2012photon, calliari2020fast}. Each of such techniques leverages the figures of merit associated to fiber monitoring: dynamic range, spatial resolution, impact on and coexistence with data transmission, acquisition time, and complexity (or cost) \cite{urban2018tutorial}.

Independently on which of these is chosen, an automated fiber measurement system (AMS) can be assembled by combining a data acquisition unit (DAU) and a signal processing unit (SPU), with the LBI figuring as one of the most promising candidates to assume the latter role. In this scenario, the figures of merit of the AMS depend both on those of the DAU and SPU, since the output result is the raw acquired data after processing; here, the output result concerns what is presented to the user by the AMS. In this context, the possibility to optimize the AMS's output, given that the figures of merit of its constituents are known, becomes an interesting avenue of research. More specifically, when the AMS is built upon a single hardware unit, such as an field-programmable gate array (FPGA), the requirements for a proposed solution become more stringent since hardware compatibility must be taken into account.

In this work, the impact of an LBI-based SPU on the AMS's output is thoroughly investigated. Since the SPU does not affect either the coexistence of the monitoring signal with data transmission or the complexity of the physical system, these are not considered. On the other hand, the spatial resolution and total measurement time are shown to be dependent on the SPU. Focus is given to identifying the characteristics of a candidate cluster that forms around the position of a fault during LBI processing. This cluster has the potential to diminish the spatial resolution of the AMS's output and even overshadow the detection of real faults. Therefore, a computationally efficient and hardware compatible compensation routine akin to deconvolution is proposed to spatially resolve the cluster. The results indicate that not only the overall precision of the SPU can be improved, but also that the raw data processing time can be reduced significantly. The latter, although seemingly not related to the cluster, is associated to alleviated requirements for the estimation routine, which, in turn, reduces the number of total iterations for the LBI algorithm.

The proposed approximate deconvolution allows achieving improved performance as far as the fiber fault detection capabilities of an LBI-based SPU go. Simultaneously, it sheds light onto the mathematical structure of the LBI algorithm by uncovering its characteristics at a regime of low number of iterations. In this regime, it is possible to extract useful information from the SPU, as shown in \cite{lunglmayr2018tim}, with good estimation performance and fast processing times; the approximate deconvolution, in turn, is shown to allow the same information to be extracted at earlier stages, representing a speed-up without loss of performance. Combined with its hardware-compatible implementation, presented and discussed in this document, the approximate deconvolution can be directly integrated to the structure presented in \cite{calliari2020jwcn} as a powerful speed-up add-on.

Moreover, this work provides the following contributions:

\begin{itemize}

    \item The observation that, when using the LBI-based sparse Kaczmarz algorithm with a limited number of iterations, the non-zero elements of an estimation vector form a cluster around the step position.
    \item The discovery  that the shape of the cluster is consistent with respect to the relative position of a fault within the fiber profile dataset and scales with the step amplitude, which enables the introduction of deconvolution methods to improve the performance of the LBI results with a limited number of iterations.
    
    \item The comparison between cluster compensation results extracted from a deconvolution filter calculated via a typically used pseudo-inverse solution and an approximate deconvolution approach.
    
    \item The demonstration that the deconvolution filter introduces additional non-zero artifacts that corrupt the solution and is, therefore, not suitable for the application at hand.
    
    \item The development of a low-complexity hardware add-on for the proposed approximate deconvolution method that, when combined with the LBI, produces accurate results even at a lower number of iterations.
\end{itemize}


The document is divided as follows. The framework of the automated fiber measurement system is laid down in Section II, with focus both on the mathematical model that describes the output of the DAU and the structure of the signal processing unit, the LBI algorithm. Section III introduces the analysis of fault clusters that are observed in the vicinities of events at lower numbers of iterations of the LBI algorithm; furthermore, the proposed approximate deconvolution cluster compensation routine is proposed and compared to a standard deconvolution method. A thorough analysis of the performance of the trend break detection routine making use of the approximate deconvolution is presented in Section IV, where significant improvements are observed allowing for the processing time to be reduced while keeping the high-quality estimation results. Section V presents the hardware structure of the approximate deconvolution in view of the current hardware implementation of the LBI algorithm as well as a complexity analysis that reveals minimum added computation effort is required to implement the proposed solution. Finally, Section VI concludes the paper and discusses future steps and points of investigation.

\section{Automated Fiber Measurement System}\label{sect:2}

The Automated Fiber Measurement System (AMS) considered here has been proposed in \cite{lunglmayr2018tim}, where a software implementation of the Linearized Bregman Iterations (LBI) algorithm was first evaluated as a potential candidate for the role of signal processing unit (SPU). Following this proposal, a hardware-compatible structure that implements the LBI algorithm has been set forth in \cite{calliari2020jwcn}, with gains in processing time reaching up to two orders of magnitude. More recently, the so-called profile-splitting methodology has been set forth \cite{amaral2020elec}, also enabling significant speed-up with no impact on the SPU's performance. Although so far the SPU and the data acquisition unit (DAU) have been presented occupying different hardware structures, simultaneous to the scientific drive to push a fully functioning AMS in a single hardware unit is the drive to optimize the features of the established AMS \cite{calliari2020jwcn}. 

The interest in an AMS stems from the necessity to perform an operation without supervision from a human user; this can be due to difficult or expensive access to the DAU, or due to the amount of operations required from the DAU and, thus, the speed in which they must be carried out. In the specific case of optical fiber monitoring, the massive optical interconnection network currently installed  worldwide, or in the process of being installed, creates the demand for periodic supervision to ensure robustness of the network. If, on one hand, a number of optical fibers are constantly being monitored, on the other, it is not necessary to supervise each monitoring round, as the probability of failure, even though not negligible, is not high. Furthermore, only in the event of a disruptive fault must a high-level (human) decision be made \cite{urban2018tutorial}. Therefore, before a SPU and a DAU are combined to produce an AMS, it is important to evaluate the reliability of each individual unit so that the requirement of human supervision can be dropped and the results can be considered robust \cite{lunglmayr2018tim}.

In the context of optical fiber monitoring, DAUs generally consist of reflectometry-based technologies, with the OTDR (with and all its variants) figuring as the most commonly employed. The output of the DAU is the so-called fiber profile, which contains information about the physical integrity of the fiber, among other characteristics. Data acquisition usually involves an optical pulse generation unit and an optical detection unit, whose characteristics determine relevant figures of merit of the DAU, such as dynamic range and spatial resolution \cite{derickson1998fiber}. Broadly speaking, the bandwidths of the pulse generation and optical detection units determine the maximum achievable spatial resolution; due to the compromise relation between spatial resolution and dynamic range, the latter is diminished as the former is improved and vice-versa \cite{derickson1998fiber}. These parameters are taken as fixed for a given DAU since, in order to modify them, different devices (e.g., detectors and associated electronics) must be employed. Considering a simplified scenario where non-linear effects associated to optical signal propagation in fibers are neglected, the DAU measures the effective impulse response of the fiber under test (FUT),
\begin{equation}
    h^{\text{eff}}\left(t\right) = h\left(t\right) * p\left(t\right),
\label{eq:conv}
\end{equation}
i.e., the convolution between the FUT's intrinsic impluse response  $h(t)$ and the transmitted pulse $p(t)$.


\begin{figure}[h]
	\centering
	\includegraphics[width=0.85\linewidth]{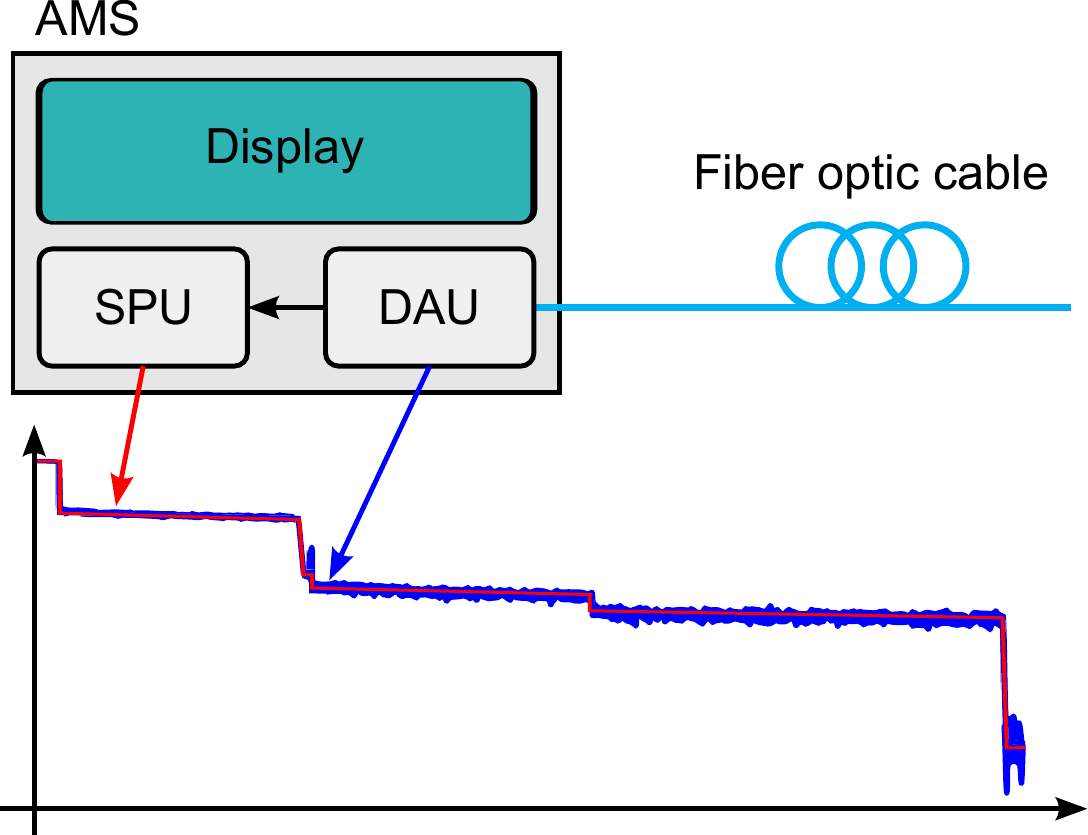}	
	\caption{Block diagram representation of the AMS containing the DAU and SPU. The raw and processed fiber profiles are depicted at the output of the DAU and SPU, respectively.}
	\label{fig:figiter}
\end{figure}

The digitized fiber profile is interpreted by the SPU, whose role is to identify the positions of trend breaks that can be associated to fiber faults \cite{lunglmayr2018tim}. Three main features of the dataset are noteworthy: the positions and magnitudes of the trend breaks are previously unknown \cite{weid2016jlt}; the fiber profile contains a linear trend, or slope, associated with the optical signal's propagation loss \cite{derickson1998fiber}; and, the dataset is corrupted by noise stemming from multiple sources \cite{lunglmayr2018tim}. Due to these, sparse signal processing techniques have found successful application in the field of automated optical fiber fault detection \cite{weid2016jlt, lunglmayr2018tim}, with the Linearized Bregman Iterations algorithm exhibiting the overall better performance demonstrated so far \cite{lunglmayr2018tim, calliari2020jwcn}. Here, the focus is on the SPU and its impact on the spatial resolution (i.e., its ability to distinguish between candidates that are close to each other) and measurement time of the AMS. This can be performed by evaluating the limitations of the SPU, which are reviewed and subsequently tested. In the following subsection, the sparse optimization problem of trend break detection in the presence of noise is presented as well as the structure of the LBI algorithm, that solves it.

\subsection{Linearized Bregman Iterations Algorithm}

The dataset output by the DAU is a vector $\bf{y}$ of length $N$, where each position contains the measured intensity within a certain spatial region associated with the DAU's detector bandwidth $\Delta \nu$. It is straightforward to observe that, given an FUT of length $X$, $N = \tfrac{4\pi Xc}{n\Delta \nu}$, where $n$ is the fiber's refractive index, $c$ is the speed of light in vacuum, the factor $2\pi$ takes into account the more general Gaussian profile, and the extra $2$ factor accounts for the round-trip time of the pulse propagation. Considering the state-of-the-art devices in the DAU, the dataset can easily reach lengths in the order of hundreds of thousands of points for medium-haul (10-20 km) optical fiber link lengths. Simultaneously, a fiber suitable for operation may not contain more than one significant fault per 5 kilometers.

Usual temporal widths of the fiber probing pulse $p(t)$ range from 1-100ns, which translates in spatial resolutions ranging from 0.1m-10m. Following this reasoning, a fiber profile, when converted to a digitized dataset, consists of about 20000 points with a conservative spatial discretization; meanwhile, the number of faults contained in the dataset is expected to be below 10. Thus, determining the positions of the unknown faults in such a dataset constitutes a sparse problem from the SPU’s perspective, since the total number of observations, $N$, dominates over the number of events, $p$, i.e., $N \gg p$ \cite{natarajan1995sparse}. Modeling fiber faults as step functions has been shown to produce accurate fault finding results \cite{lunglmayr2018tim}; for such, it is necessary to introduce the candidate matrix (or dictionary) A, which contains step functions in its columns.
Due to the intrinsic attenuation of light as it propagates through optical fibers, the matrix $\mathbf{A}$ $(\text{dim}(\mathbf{A})=N \times (N+1))$  must also account for a negative slope, as follows:

\begin{equation}
		\mathbf{A}=
		\begin{bmatrix}
			
			1 & 1 & 0 & 0 & \dots & 0 & 0 \\
			2 & 1 & 1 & 0 & \dots & 0 & 0 \\
			3 & 1 & 1 & 1 & \dots & 0 & 0 \\
			\vdots & \vdots & \vdots & \vdots & \ddots & \vdots & \vdots \\
			N-1 & 1 & 1 & 1 & \dots & 1 & 0 \\
			N & 1 & 1 & 1 & \dots & 1 & 1 \\
		\end{bmatrix}
	\end{equation}
Finally, the signal model is given by:
\begin{equation}
    \mathbf{y}=\mathbf{A}\betab, 
\end{equation}
where $\betab$ is a $(N+1) \times 1$ sparse vector and its nonzero components index the positions of the faults in the profile. In order to estimate the parameter $\betab$, one can solved the following $\ell_1/\ell_2$ optimization problem, which provides a sparse solution $\hat{\betab}$:

\begin{equation}
\hat{\betab}=\arg \min_{\betab} \lambda||\betab||_1 + \tfrac{1}{2}||\betab||_2^2 \;  \text{s.t.} \; \bf{A}\betab = \bf{y}.
	\label{eq:l1/l2}
\end{equation}

By including the term $||\betab||_2^2$, the objective function becomes strongly convex, fostering efficient solution algorithms. In other words, the $\ell_2$-norm is added to the cost function (instead of only using the $\ell_1$-norm) to enable efficient algorithms to solve the posed problem while ensuring sparsity by choosing a large enough value for $\lambda$. One of such efficient algorithms is the Linearized Bregman iterations \cite{LB_theory,LBdual,bregman_osher}, which has been proven to converge to the solution of problem \eqref{eq:l1/l2}. The LBI algorithm is a linearized version of the Bregman algorithm, which, in turn, is based on successively minimizing the so-called Bregman distance. This algorithm has been shown to yield a sparse solution of (4) even in the presence of noise \cite{dirk_lorenz}.


The main advantage of the LBI is that the minimization problem has a closed form solution. Moreover, different versions of the LBI, based on its original formulation, were proposed in \cite {lunglmayr2018tim, calliari2020jwcn, amaral2020elec} for fiber fault detection, which introduced different strategies for efficient digital hardware implementation. More specifically, the Kaczmarz variant \cite{dirk_lorenz, michael_kaczmarz},  which yields the simple structure of Algorithm \ref{alg:LB}, proven to be very efficient and, thus, used throughout this work. Algorithm \ref{alg:LB} is an implementation of the LBI method based on the sparse Kaczmarz variant which instead of using the whole matrix A in each iteration sweeps through the rows of $\mathbf{A}$ in a cyclic manner. The shrink function in line 7 is defined as follows:

\begin{equation}
\text{shrink}(\mathbf{x}, \lambda)= \max(\mathbf{x}-\lambda, 0)\cdot \text{sign}(\mathbf{x}),
\end{equation}
the shrink function introduces sparsity due to the fact that it sets all values of the argument $\mathbf{x}$ to zero that have absolute values smaller than $\lambda$.

For simplicity, the number of iterations of the LBI algorithm will henceforth refer to $\alpha$ in Alg. \ref{alg:LB} (also known as the \textit{number of iterations per sample} \cite{calliari2020jwcn}), i.e., a full algorithm loop over all $N$ positions of the vector $\bf{y}$.

\begin{algorithm}
	\caption{Linearized Bregman-based Sparse Kaczmarz \label{alg:LB}}
	\textbf{Input:}
	$\mathbf{A},\mathbf{y}$, $\alpha$, $N$
	
	\textbf{Initialization:}
	$\hat{\mathbf{\betab}}^{0}=\mathbf{0}$, $\mathbf{v}^{0}=\mathbf{0}$, $L=\alpha N$, $\lambda=0.5$ 	
	\medskip
	
	\For{$i=1:L $}{
	  $k \leftarrow ((i-1) \mod N)+1$
	  
	  $\mathbf{v}^{i+1}=\mathbf{v}^{i}+ \frac{1}{\left\| \mathbf{a}_k \right\|_2^2} \mathbf{a}_k (y_k -\mathbf{a}_k^T \betab^i) $
	  
	  \For{$j=1: k $}{
	  $\hat{{\beta}}_j^{i}=  \text{shrink} (v_j^{i},\lambda)$
	  }
	 
	}	
	\textbf{Output:}
	$\hat{\mathbf{\betab}}$
\end{algorithm}

\section{Fault Clusters}\label{sect:motivation}

Despite the unmatched performance and time-efficiency of the LBI -- when applied to fiber fault detection --, the compromise between these two features is difficult to gauge in a specific application or scenario. Most of the results are drawn based on a large testbench containing several different datasets that exhibit distinct characteristics. To summarize the statistical analysis, different figures of merit can be computed, with the Matthew's Correlation Coefficient (MCC) \cite{matthews1975comparison} -- which can be calculated based on a Contingency Table -- figuring as a robust choice since it provides an informative and truthful metric in evaluating binary classifications; a high score is produced if the estimation succeeds in identifying the underlying model of the datasets (in this case the positions and magnitudes of the faults). Throughout the analysis presented here, the testbench of datasets are simulated using the protocol described in \cite{lunglmayr2018tim}, with 5 events per dataset with magnitudes ranging between 0.1 and 5 dB. The Contingency Table, in turn, is a widely used tool for statistical analysis of binary classification processes and displays the (multivariate) frequency distribution of the associated variables. Most relevant to the present analysis is the fact that the MCC is the only binary classification rate that generates a high score only if the binary predictor was able to correctly predict a large number of positive data instances and a large number of negative data instances.

\begin{figure}[h]
	\centering
	\includegraphics[width=1\columnwidth]{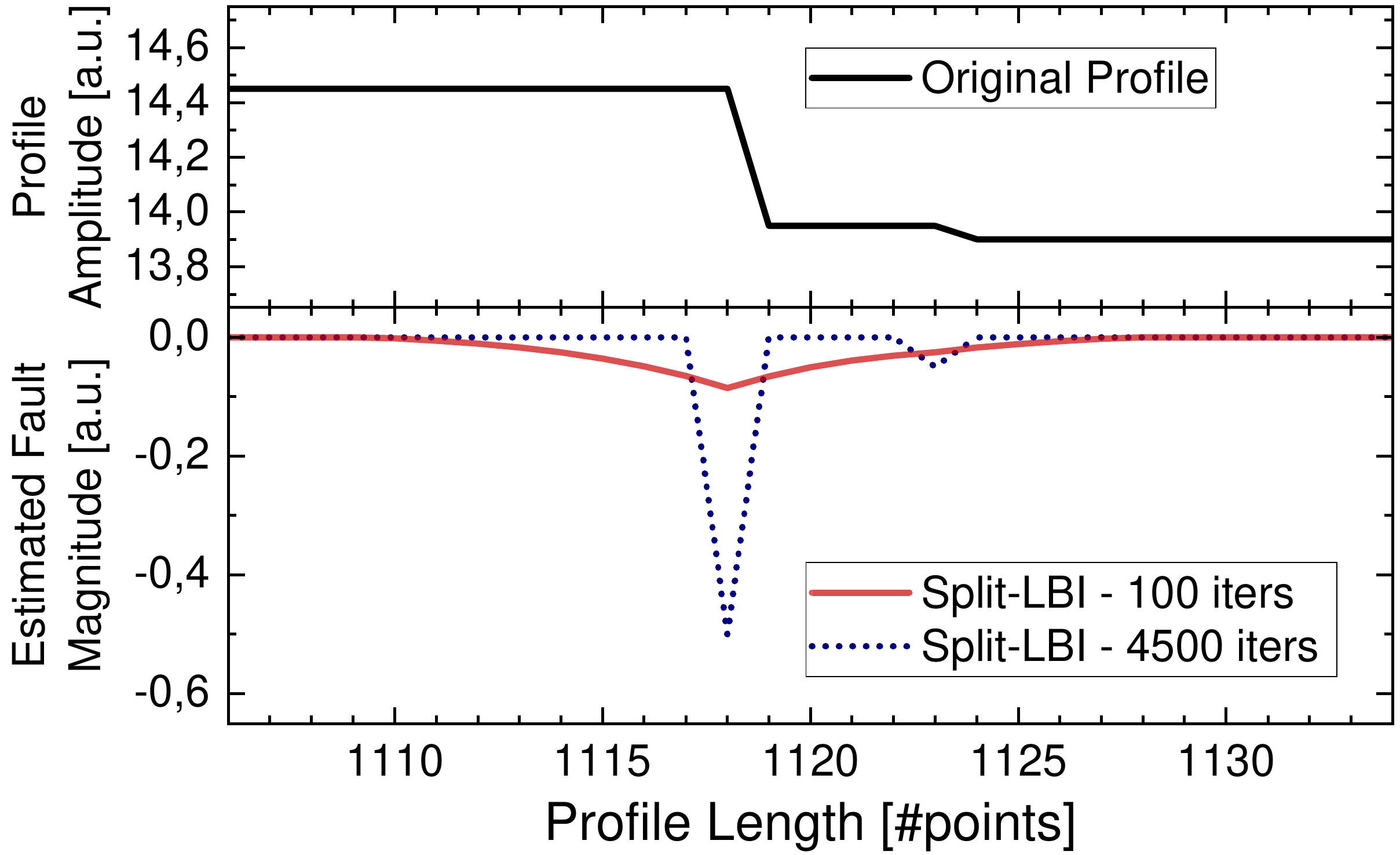}
	\caption{Performance of the Split-Profile LBI algorithm for different number of iterations for a scenario with a noise-free profile and two close faults. The fault cluster clearly appears when the number of iterations is lower.}
	\label{fig:clusterEffect}
\end{figure}

The advantages of a simulated testbench are the possibility to control the level of noise in the system, and the knowledge of the true underlying model: the former follows from the analysis of the noise present in the real-world experimental dataset \cite{lunglmayr2018tim}; and the latter allows extracting the Contingency Table and, thus, the MCC. The disadvantage is that, as previously mentioned, peculiar characteristics of individual datasets may be overlooked due to the large number of datasets in the testbench that enable a statistically relevant analysis of the performance of the algorithm to be drawn. Upon inspection of the LBI's estimated output $\hat{\boldsymbol{\beta}}_{\text{raw}}$ of individual datasets, one of such peculiarities stands out as recurring. Consider the results depicted in Fig. \ref{fig:clusterEffect}: here, a noiseless dataset was created with two events purposefully close to each other ($p_1-p_2=\Delta=5$ points). Due to the magnitude difference and proximity of the events, the two faults are not distinguishable for a low number of iterations, but become discernible if the algorithm's elapsed time is significantly longer.

Noteworthy from the result of Fig. \ref{fig:clusterEffect} is the fact that the fault event of highest magnitude overshadows its lowest neighbor due to what is herewith dubbed a \textbf{fault cluster}: the spread of the localization of the event around its true position or, formally, the fact that positions of the estimated vector $\hat{\boldsymbol{\beta}}_{\text{raw}}$ in the vicinity of the fault are different from zero for a certain range of number of iterations. The cluster can be analyzed from the same perspective of Eq. \ref{eq:conv}, i.e., that of the reduction of spatial resolution in the DAU due to the time duration of the probing pulse; the difference, here, is that the observed fault cluster is a completely algorithmic effect and has no ties to the physical limitations of the DAU. This analysis leads to the conjecture that the spatial resolution of the AMS is, in fact, limited by:
\begin{equation}
    h^{\text{AMS}}\left[k\right] = h\left[k\right] * p\left[k\right] * C\left[k\right],
\label{eq:conv_ck}
\end{equation}
where $k$ is a discrete variable that indexes the digitized fiber profile and $C\left[k\right]$ represents the cluster shape, which is expected to impact the result through a convolution. In order to confirm the analysis, it is imperative that the characteristics of the cluster are well understood: in other words, whether $C\left[k\right]$ changes with the number of iterations, and position and magnitude of a fault.

\begin{figure}[h]
	\centering
	\includegraphics[width=1\linewidth]{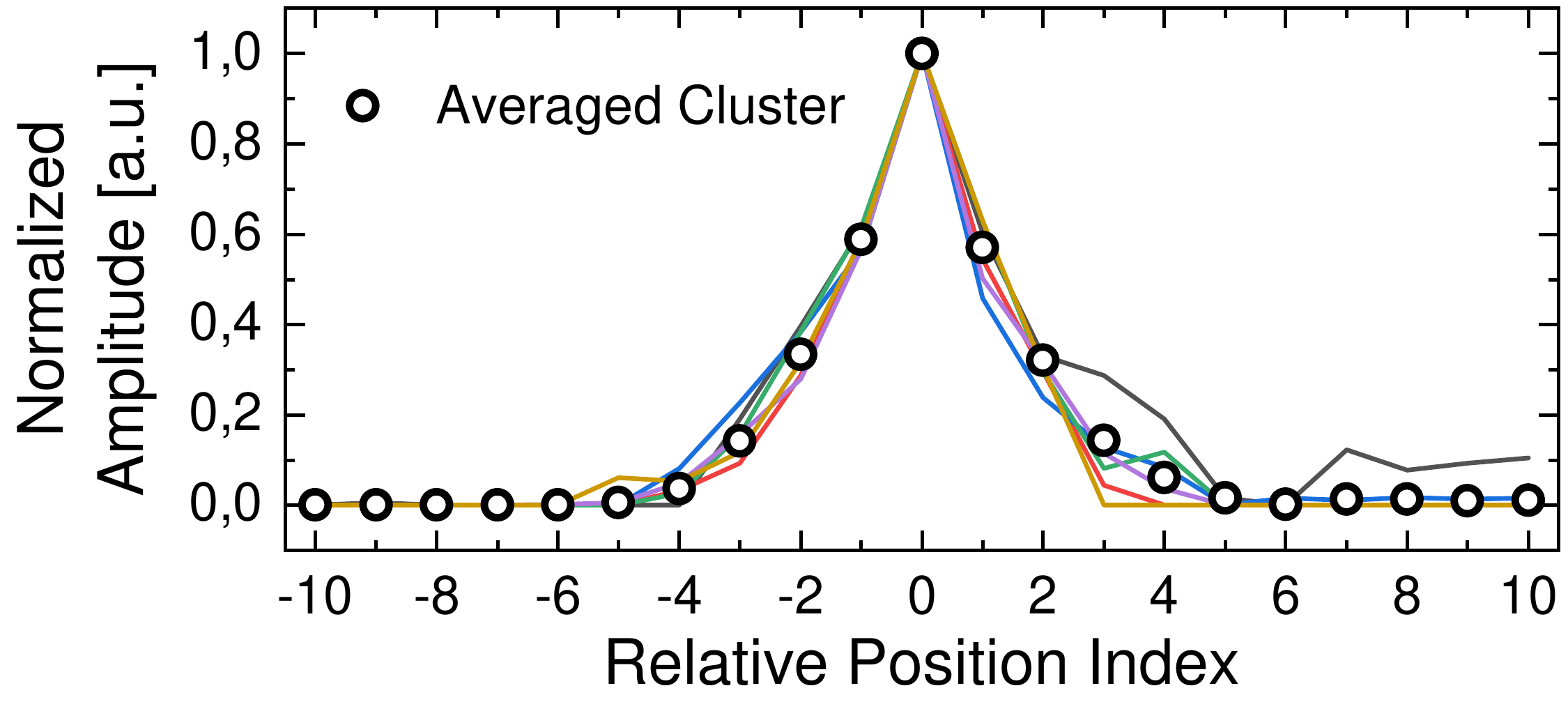}
	\caption{Result of the normalization, superimposition, and averaging of several fault clusters. Although only a few fault cluster shapes are depicted, over 100 noiseless profiles were processed with faults at different relative positions along the dataset to determine the averaged cluster shape that is used throughout the analysis. The solid lines represent the different normalized clusters shapes extracted from the simulated profile testbench.}
	\label{fig:clusterShape}
\end{figure}

From Fig. \ref{fig:clusterEffect}, it is already possible to ensure that $C\left[k\right]$ changes with the number of iterations, which is expected: with a high enough number of iterations, the algorithm's response tends to an impulse response ($C\left[k\right] \rightarrow \delta\left[k\right]$) and the AMS recovers the physically-limited resolution of the DAU; nevertheless, it is impractical to process several long datasets with the number of iterations required so the above condition is reached. On the other hand, with sufficient knowledge of the cluster shape, it is possible to revert its effect through a deconvolution process, provided that the shape of $C\left[k\right]$ is constant or slowly varying for different fault positions and magnitudes, which is what is shown in Fig. \ref{fig:clusterShape}. For these results, over 100 different profiles (only a few are actually shown in Fig. \ref{fig:clusterShape}) were processed by the split-profile LBI algorithm \cite{amaral2020elec} (350 iterations, 4500-points split-profile length); the positions of the faults were randomized as well as their magnitudes and, for visualization purposes, the results were normalized, superimposed, and averaged.

It becomes clear that the shape of the cluster is somewhat constant (or at least slowly varying as a function of position and magnitude) such that it can, for a fixed number of iterations, be reliably and reproducibly used in a compensation procedure to increase the spatial resolution even in a regime such that $C\left[k\right] \neq \delta\left[k\right]$. It is important to point out that, in the analysis that follows, the interest is to compensate the effect of the algorithm on an ideal (noiseless) dataset, removing the negative contribution of $C\left[k\right]$. In other words, the study of the cluster analysis is performed in a controlled ideal scenario so that it can be subsequently applied to datasets containing the noise contribution derived from the experimental data model \cite{lunglmayr2018tim}. Following this procedure, compensation of any effect that is not solely due to the algorithm's performance is avoided. This has a generalizing effect, i.e., any dataset contaminated by noise can be processed by the same compensation procedure since it targets the algorithm's effect rather than that of the noise.

\subsection{Deconvolution Filter and Approximate Deconvolution}

If one takes the assumption formally stated in Eq. \ref{eq:conv_ck} to be true, the peak-broadening effect observed at the output $\hat{\boldsymbol{\beta}}_{\text{raw}}$ of the LBI algorithm is due to the convolution of the physically-limited DAU's output given by Eq. \ref{eq:conv} and the cluster shape $C\left[k\right]$. In order to compensate this effect, one could, then, resort to a deconvolution filter, i.e., a signal $g\left[m\right] \in \mathbb{R}^ M$ that is applied to $\hat{\boldsymbol{\beta}}_{\text{raw}}$ with the following property:
\begin{equation}
	\text{C}[k]*\text{g}[m]=\delta[m-k],
	\label{eq:de_convol}
\end{equation}
where the vector $\mathbf{g}$ corresponds to an FIR filter of length $m$ that eliminates the effect of $C\left[k\right]$ generating a new estimation vector $\hat{\boldsymbol{\beta}}_{\text{deconv}}$. Following this rationale, it would be possible to detect the presence of faults in $\hat{\boldsymbol{\beta}}_{\text{deconv}}$ otherwise indiscernible in $\hat{\boldsymbol{\beta}}$, thereby extracting more information about the true underlying model even in a regime of low number of iterations. The most reliable way of estimating the vector $\mathbf{g}$, in possession of the averaged cluster shape (refer to Fig. \ref{fig:clusterShape}), is through a least squares optimization, i.e.,
\begin{equation}
    \mathbf{g}= (\mathbf{D}^T \mathbf{D})^{-1} \mathbf{D}^T \boldsymbol{\delta}[m-k],
\end{equation}
where the convolution matrix $\mathbf{D} \in \mathbb{R}^{(K+M-1) \times M}$ is defined such that $\mathbf{D}\mathbf{g} = \text{C}\left[k\right]*\text{g}\left[m\right]$.

Although mathematically rigorous, the above procedure can become taxing in terms of complexity and, especially, in terms of hardware integration \cite{calliari2020jwcn}. At this point, it is useful to address a feature of the fault detection procedure detailed in \cite{lunglmayr2018tim, amaral2020elec}: a so-called \textit{peak-detection} routine is applied to the raw data output from the LBI processing core (Algorithm \ref{alg:LB}) before it is presented to the user. Peak-detection is a low-complexity routine that identifies positions in the vector $\hat{\boldsymbol{\beta}}_{\text{raw}}$ that exhibit higher magnitudes than both its neighbours, thereby being associated to peaks and, thus, to detectable events. Since the loss in spatial resolution due to the cluster has been shown to take place when a fault with high magnitude overshadows a neighbouring smaller fault, cluster compensation is only necessary in the vicinities of detected peaks \textit{after} the peak-detection routine is applied to $\hat{\boldsymbol{\beta}}_{\text{raw}}$. This observation, combined with the knowledge of the shape of the cluster, has the potential to diminish the complexity associated to the cluster compensation procedure.

\begin{figure}[h]
	\centering
	\includegraphics[width=0.95\columnwidth]{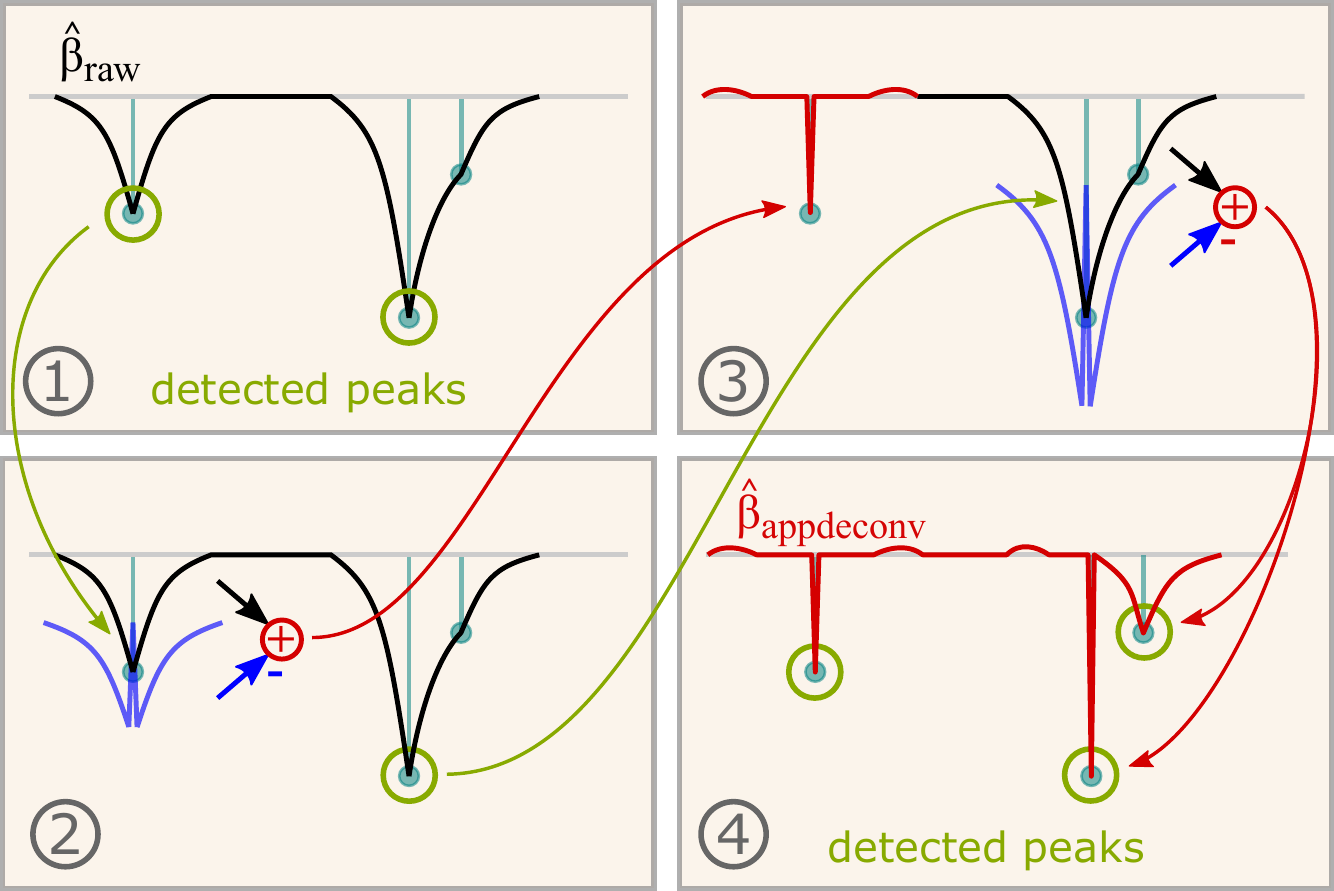}
	\caption{The proposed approximate deconvolution routine in schematic representation with its three phases: (1) peak-detection; (2-3) local cluster compensation; and (4), final peak-detection. Green arrows represent the memory of the initially detected peaks, while red arrows represent the application of the approximate deconvolution procedure.}
	\label{fig:approxdeconv}
\end{figure}

Fig.~\ref{fig:approxdeconv} schematically describes the principle of the herewith dubbed \textbf{approximate deconvolution} method. First, the raw output of the LBI core is subjected to peak-detection (1); the detected peaks in phase (1) are, then, processed successively. For each peak, a static compensation function is scaled depending on the detected peak's magnitude and subtracted element-wise from $\hat{\boldsymbol{\beta}}_{\text{raw}}$ (2-3). The static compensation function takes the form of the $C\left[k\right]$ but with its central position (the one associated to the peak) set to zero. Because the central position of the compensation function is zero, the magnitude of the detected peak in (1) is preserved, whereas the neighbouring positions have their cluster contributions eliminated. After processing of the identified peaks (2-3), the compensated signal $\hat{\boldsymbol{\beta}}_{\text{appdeconv}}$ is once again subjected to the peak-detection routine (4), so that those peaks overshadowed by the cluster shape (if any) can be identified.

\section{Performance Analysis}

\begin{figure}[h]
	\centering
	\includegraphics[width=1\linewidth]{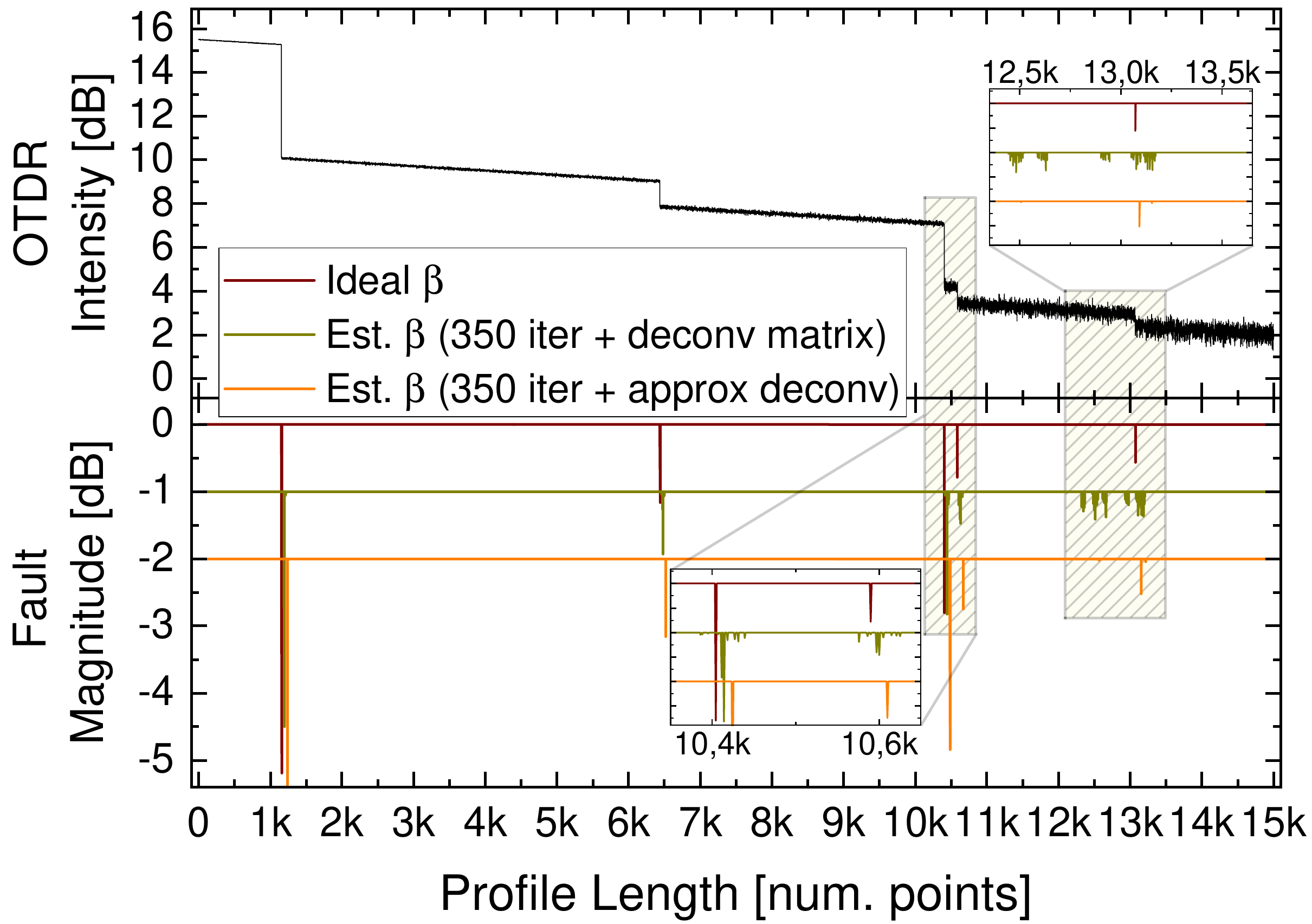}
	\caption{Performance of the Split-Profile LBI algorithm after 350 iterations and application of the deconvolution and approximate deconvolution methods. The lower panel data series that presents the ideal and estimated $\beta$ values for each position (as well as the inset data) are offset both vertically and horizontally so that the results can be visualized.}
	\label{fig:figconvMatrix}
\end{figure}

The performance of both methods (deconvolution and approximate deconvolution) were initially evaluated and compared in a simplified controlled scenario: a single dataset with faults purposefully spread around to avoid the cluster-induced loss of spatial resolution; this allows testing the overall performance of the methods for a more general dataset in the presence of noise. The results, depicted in Fig. \ref{fig:figconvMatrix}, reveal an interesting behavior from the deconvolution method: the oscillatory characteristic of the FIR filter $\mathbf{g}$ floods the estimation output with so-called false positives. From a system-theoretical point of view these oscillations stem from the errors introduced by the least squares optimization (Eq. \ref{eq:de_convol}) when attempting to estimate the inverse of an FIR filter with yet another FIR filter. As can be observed, this effect is even more pronounced in presence of noise, which corresponds to the real-world conditions simulated in the profile of Fig. \ref{fig:figconvMatrix}. Conversely, the approximate deconvolution method is able to perfectly extract the underlying model even in presence of noise, advocating for its use not only from the point of view of complexity reduction in the procedure, but also in terms of quality and consistency of the estimation results.

As discussed so far, the approximate deconvolution, that derives from the analysis of the fault cluster, attempts to extract information from the true underlying model of a dataset with minimum added complexity even at a regime where the low number of iterations masks the presence of certain faults. Increased performance due to the application of this method would be evidenced by an increase in the computed score of the MCC when processing a testbench containing a large number of datasets. On the other hand, it is also imperative to test the procedure on a real-world experimentally acquired dataset to avoid restricting the analysis to simulated datasets. In this section, both analyses are presented, where a fine-tuning of the parameters is initially performed using the statistical analysis enabled by a simulated testbench.

The sole parameter that requires fine-tuning is the length of the approximate deconvolution vector that is subtracted element-wise from the raw output of the LBI. Although longer lengths will likely lead to better estimations, they also lead to increased complexity in the procedure; for that reason, it is important to strike a good balance between these two figures of merit. Furthermore, and based on the analysis of the previous section, the coefficients are pre-calculated since they need only be scaled by the peak magnitude: the shape of the compensation function is static. In a hardware implementation, discussed in detail in Section \ref{sect:hardware}, a static ROM memory can be used to store the cluster compensation vector coefficients, which represents yet another motivation to keep its length as short as possible without compromising the performance.

\begin{figure}[h]
	\centering
	\includegraphics[width=1\linewidth]{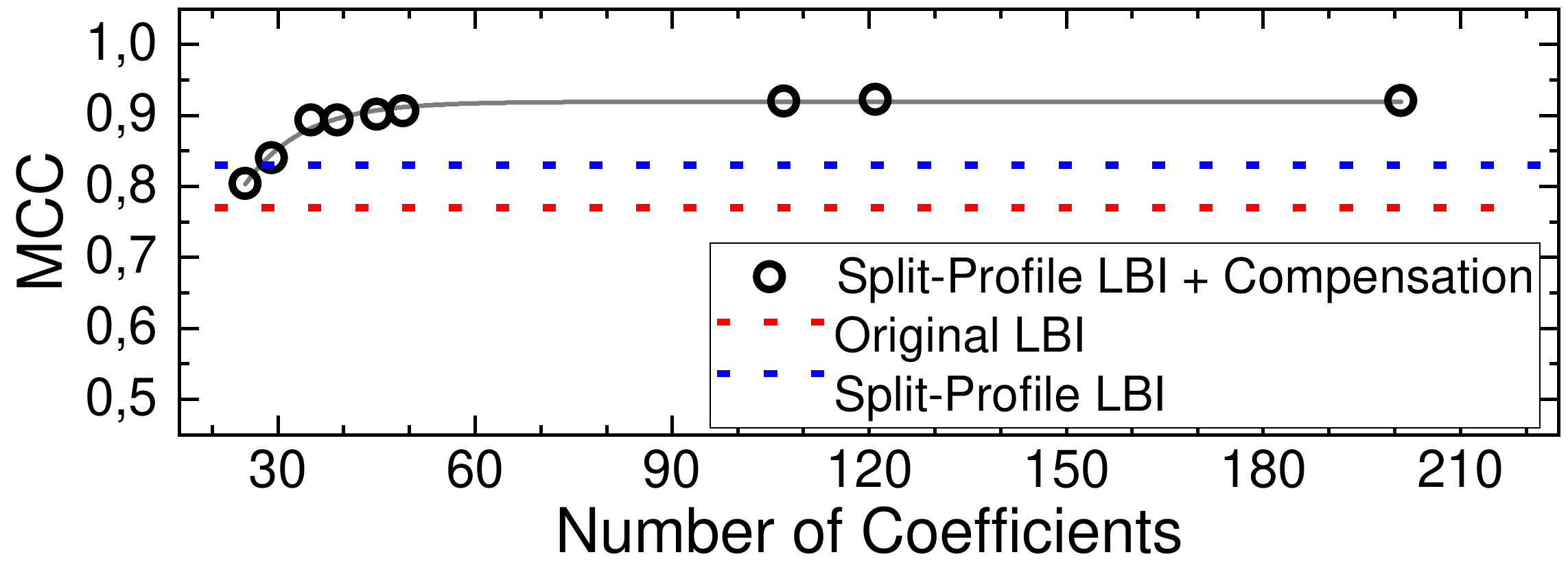}
	\caption{Performance of the Split-Profile LBI algorithm after application of the approximate deconvolution for different number of coefficients. The MCC score for the original and split-profile LBI are constant since the number of coefficients does not impact their estimation result.}
	\label{fig:coeffsMCCanalysis}
\end{figure}

In Fig. \ref{fig:coeffsMCCanalysis} the computed value of the MCC is depicted as a function of the number of coefficients for a testbench containing 100 15000-points-long datasets, where the iterations were fixed at 350 and the split-profile length at 4500. To provide a reference of the increased performance enabled by the proposed cluster compensation routine, the MCC values of the original and split-profile versions of the LBI algorithm are also presented. As expected, by increasing the number of coefficients, the performance of the estimation increases until it ends up saturating; from the results, an optimal compromise between length and performance has been set to 65 coefficients. Surprisingly, the most remarkable feature of this result is not the trend of the performance as a function of the number of coefficients employed, but, rather, how much the performance increases: from an average of 0.83 to an astounding 0.92, a value that had not been previously observed even for higher numbers of iterations \cite{calliari2020jwcn}.

\begin{figure}[h]
	\centering
	\includegraphics[width=1\linewidth]{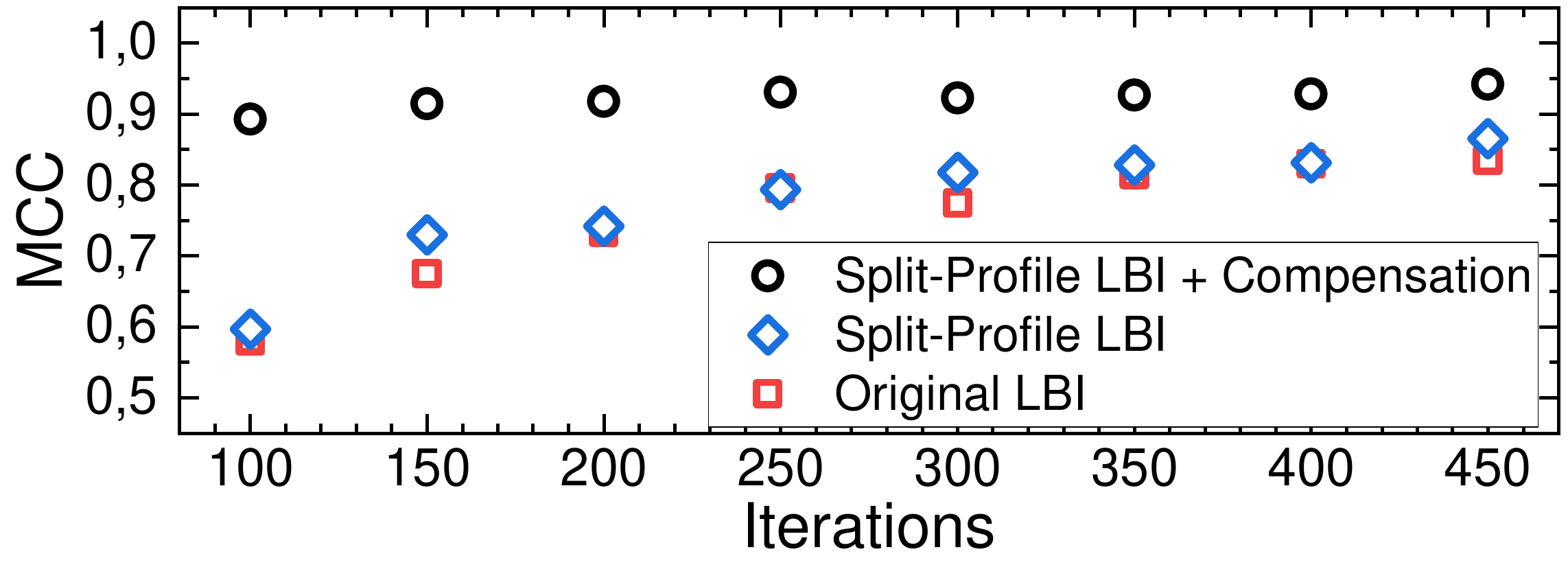}
	\caption{Performance of the Split-Profile LBI algorithm after application of the approximate deconvolution for different number of iterations and a fixed compensation vector length (65). Although the gains for lower number of iterations is more prominent, the performance boost extends also to higher numbers evidencing the positive impact of the proposed compensation routine.}
	\label{fig:itersMCCanalysis}
\end{figure}

This observation begs the question: is it possible to reduce the number of iterations even further while maintaining the same performance due to the cluster compensation routine? To answer it, the results of Fig. \ref{fig:itersMCCanalysis} are extracted, where the number of coefficients are set to 65, but the number of iterations is varied between 100 and 450 for the same testbench used in Fig. \ref{fig:coeffsMCCanalysis}. A tremendous performance boost is observed for as low as 100 iterations, which holds up for all values of the number of iterations until 450. In order to provide an insightful analysis of these results, it is interesting to recall the MCC in mathematical form:
\begin{equation}
\small{\text{MCC}}= \small{\frac{TP \cdot TN -FP\cdot FN}{\sqrt{(TP\!+\!FP)(TP\!+\!FN)(TN\!+\!FP)(TN\!+\!FN)}}},
\end{equation}
where TP, TN, FP, and FN stand for true-positives, true-negatives, false-positives, and false-negatives, respectively. These values can be directly extracted from a Contingency Table and, to maintain the strictest level of performance evaluation, only faults identified at matching positions with respect to the underlying model are considered true-positives.

\begin{figure}[h]
	\centering
	\includegraphics[width=1\linewidth]{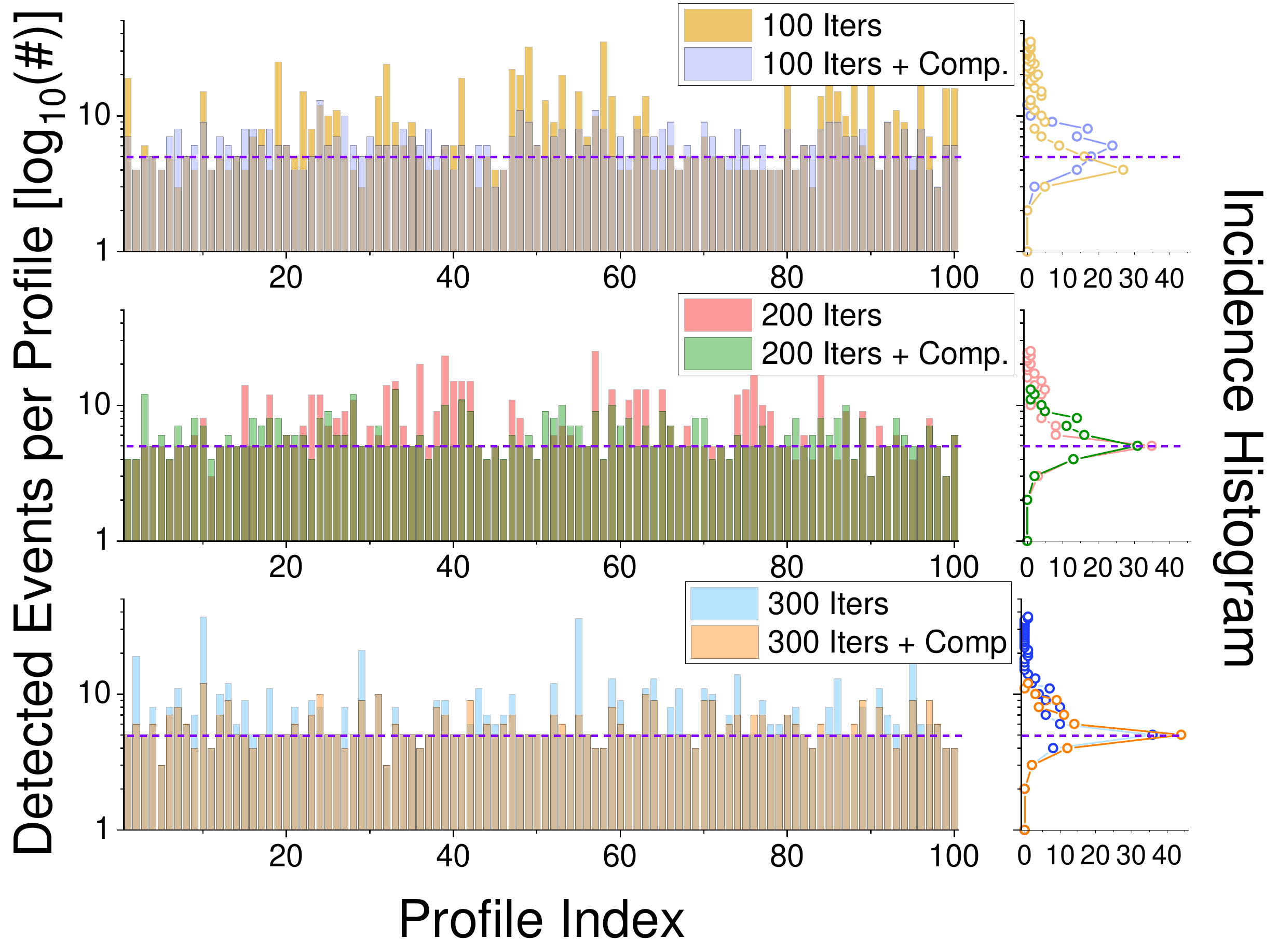}
	\caption{Detailed analysis of the incidence of identified events over a testbench of 100 simulated profiles for different numbers of iterations with and without the compensation routine. The effect of the approximate deconvolution is a simultaneous increase in values of TP and decrease in values of FP, leading to a performance boost even for lower number of iterations of the split-profile LBI algorithm.}
	\label{fig:PeakIncidence}
\end{figure}

As already demonstrated in \cite{lunglmayr2018tim}, due to the sparse characteristic of the trend break detection problem in optical fiber profiles, the values of TP and FP have a much higher impact on the performance of the algorithm than those of FN and TN, which are intrinsically low and high, respectively, due to the combined $\ell_1/\ell_2$ minimization. Although an increase in TP values is expected from the cluster compensation routine (since faults otherwise indiscernible are revealed), this impact alone would not be statistically sufficient to improve the MCC to the level observed in Fig. \ref{fig:itersMCCanalysis}, which indicates that the approximate deconvolution is also filtering out false positives. This can, indeed, be observed from the results of Fig. \ref{fig:PeakIncidence}, where the numbers of events identified by the split-profile LBI algorithm, with and without compensation using the proposed approximate deconvolution, are depicted.

If, on one hand, there is a clear increase in the identified number of events up to 5 in the histograms due to the proposed compensation routine (the number of actual events in the underlying model), there is also a decrease in the events above 5, which corresponds to a decrease in the value of FP. Based on the results of Figs. \ref{fig:coeffsMCCanalysis}--\ref{fig:PeakIncidence}, 200 iterations, 65 coefficients, and a split-profile length of 4500 are henceforth set as nominal parameters of the algorithm due to the good compromise between performance and complexity of the algorithm. This update represents a higher than two-fold speed-up factor with respect to previous recent results \cite{amaral2020elec} and a performance boost that translates into an increase of 16\% on the averaged MCC score. The processing times are reduced because the approximate deconvolution add-on allows to maintaining (or even increasing) the performance level at a reduced number of iterations (which is the main factor defining the processing time) while the complexity of the approximate deconvolution can be considered negligible in comparison. As we observed in the results of Fig. \ref{fig:itersMCCanalysis}, an MCC of 0.9 (if sufficient for a given use case) can be achieved with only 100 iterations per sample in comparison with former results where 450 iterations per sample were needed. This considerable reduction of approximately 80 \% of the iterations translates into an over four-fold reduction of the processing time.

It is important to observe that, depending on the application-dependent sought-after accuracy of detection or processing speed-up, the number of iterations can be further manipulated. As an example, it is possible to bring the number of iterations down to 100 and, in a hardware implementation, process a 20000-long dataset in a medium-sized FPGA such as the \textsc{Stratix V} in under 2 seconds \cite{amaral2020elec} while still having access to an averaged MCC above $0.85$.

\section{Hardware Implementation and Complexity Analysis} \label{sect:hardware}

Efficient hardware implementation of the proposed approximate deconvolution compensation routine must satisfy two conditions. First, that the number of clock cycles dedicated to the compensation routine does not overcome the number of clock cycles that would be necessary for the LBI algorithm to reach the same performance; following the results of Fig. \ref{fig:itersMCCanalysis}, the LBI would require at least 450 iterations to approximate the performance reached by the compensation routine after 200 iterations. Second, that the proposed hardware is compatible with the hardware structure developed for the LBI algorithm \cite{calliari2020jwcn}. For the latter point, and also in the context of the subsequent discussion, it is important to recall that the LBI hardware, even though implemented with a bank of parallel Block RAMs, is capable of flushing the values of the stored vector $\hat{\boldsymbol{\beta}}_{\text{raw}}$ sequentially after the iterations elapse.

\begin{figure}[h]
	\centering
	\includegraphics[width=1\linewidth]{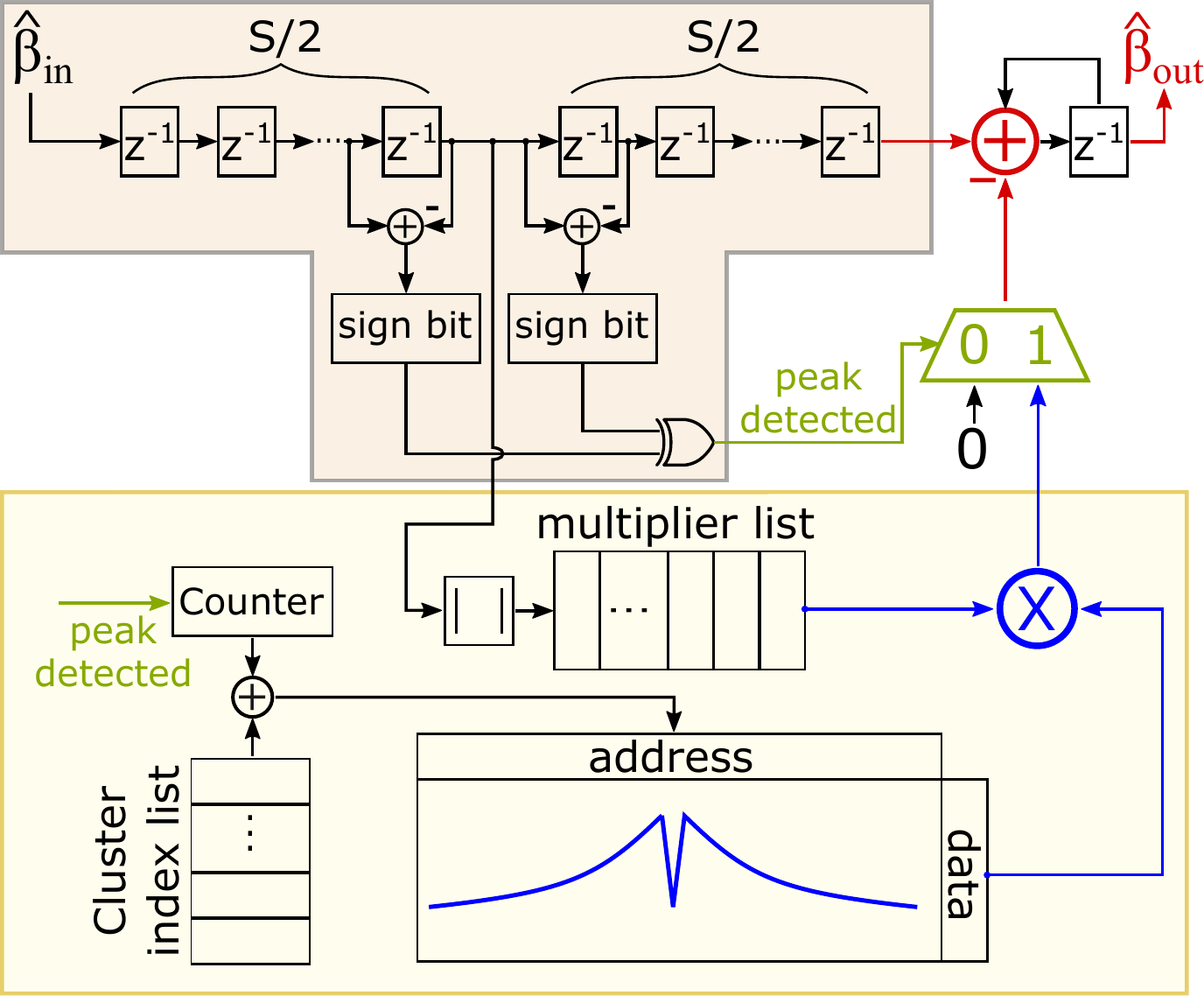}
	\caption{Proposed hardware implementation of the peak detection and compensation routines. Note that the color scheme follows that of Fig. \ref{fig:approxdeconv}. Sequential access to the values of $\hat{\boldsymbol{\beta}}_{\text{raw}}$ is ensured by the hardware structure of the LBI algorithm presented in \cite{calliari2020jwcn}.}
	\label{fig:deconvhw}
\end{figure}

The proposed hardware consists of an amalgamation of the peak-detection unit, that consists of a shift register and simple arithmetic operations, and the actual compensation arithmetic unit, as depicted in block diagram format in Fig. \ref{fig:deconvhw}. In order to adapt the compensation arithmetic, the length $s$ of the shift register is implemented to match the length of the compensation vector, whose normalized coefficients are stored in a static ROM. The peak detection arithmetic takes place at index $s/2$ of the shift register such that, in case a peak is detected, compensation of the positions at the edge of the cluster ($s/2$ away from the peak at the center) can take place as the corresponding positions of $\hat{\boldsymbol{\beta}}_{\text{raw}}$ exit the shift register unit. The peak detection routine is a simple nearest neighbour check, which must be followed by storage of the corresponding peak magnitude (in the \textit{multiplier list} of Fig. \ref{fig:deconvhw}) so that the cluster shape can be correctly scaled. The detection of a peak also triggers a counter that acts on the static memory that stores the compensation coefficients and runs from 1 to $s$.

The compensation routine depicted also contains a self-adjusting mechanism, whereby events that fall within each other's cluster range can be compensated accordingly. For that reason, a cluster index list is adopted such that a new detected peak position can be stored until its cluster contribution has been removed from the estimation vector. In order to allow for multiple clusters to be simultaneously compensated, the master clock that controls the shift register is disabled and a \textit{sum and accumulate} structure (upper right corner of Fig. \ref{fig:deconvhw}) is activated until the cluster index list is completely scanned. The presented structure takes as many clock cycles to correct an instance of $\hat{\boldsymbol{\beta}}_{\text{raw}}$ as peaks are found; in the worst possible case, the $p$ peaks all fall within each other's cluster region (i.e., their distances are all smaller than s/2), requiring the shift register to halt multiple times. The number of clock cycles required to correct one entry of $\hat{\boldsymbol{\beta}}_{\text{raw}}$ corresponds to the current length of the cluster index list, which is at most $p$ at full overlap; before the full overlap it decays symmetrically to $p-1$, $p-2$, and so on. Therefore, the summation of all these contributions leads to a total number of extra clock cycles -- in the worst possible case --  equal to the product $sp$. According to the results of Fig. \ref{fig:PeakIncidence}, a N=15000 profile is likely to produce, after 200 iterations, a number of detected peaks in the range 10-20. Taking the higher value, the total number of clock cycles required by the structure of Fig. \ref{fig:deconvhw} to process ($\hat{\boldsymbol{\beta}}_{\text{raw}} \rightarrow \hat{\boldsymbol{\beta}}_{\text{appdeconv}}$) an $N=15000$ output is merely $CC=16200$. Compared to the number of clock cycles required for the LBI itself, typically in the tens or hundreds of millions, the processing time required by the approximate deconvolution, albeit its disruptive impact in the performance of the estimation, is considered negligible.

\section{Conclusion}

The correct detection of faults in optical fibers plays an important role in the design of robust optical networks, contributing to a better user experience and optimization of resources on the operator side. One of the defining factors of fault detection is spatial resolution, the limitation of which is a contribution from data acquisition and processing in an automated fiber measurement system. In this paper, the spatial resolution limitation induced by the Linearized Bregman iterations algorithm -- adapted to trend break detection -- has been thoroughly investigated. The observation that a limited finite number of iterations of the algorithm yields so-called detection clusters in the true event's neighbouring positions allowed for an efficient compensation routine to be proposed and tested. The effect of such compensation has been shown to extend beyond increased spatial resolution, boosting the estimation performance of the algorithm while allowing for reduced processing times. The associated hardware structure that implements the approximate deconvolution is readily adaptable to the existing LBI's hardware. These results pave the way for a near-term fully functional high-performance, efficient, and embedded AMS, which figures as the main future step of this research.

\section*{Acknowledgment}

The authors would like to thank the Austrian COMET. This work is supported by: the COMET-K2 “Center for Symbiotic Mechatronics” of the Linz Center of Mechatronics (LCM), funded by the
Austrian federal government and the federal state of Upper Austria. The authors would like to acknowledge the support from Brazilian agencies CNPq, Capes, and FAPERJ.

\vspace{0.82cm}
\bibliographystyle{IEEEtran}
\bibliography{References}

\end{document}